\documentclass[journal]{IEEEtran}

\usepackage{amsmath,amssymb,amsfonts}
\usepackage{algorithmic}
\usepackage{graphicx}

\usepackage{multirow}
\usepackage{subcaption}
\usepackage{caption}
\usepackage{float}
\usepackage{xcolor}
\usepackage{booktabs}
\usepackage{hyperref}

\begin{document}




\title{Systematic Clinical Evaluation of A Deep Learning Method for Medical Image Segmentation: Radiosurgery Application}


\author{
Boris Shirokikh,  
Alexandra Dalechina,
Alexey Shevtsov,
Egor Krivov,
Valery Kostjuchenko,
Amayak Durgaryan,
Mikhail Galkin,
Andrey Golanov,
and Mikhail Belyaev  

\thanks{The results have been obtained under the support of the Russian Foundation for Basic Research grant 18-29-26030. (Corresponding authors: B. Shirokikh; A. Dalechina; M. Belyaev.)}

\thanks{B. Shirokikh, A. Shevtsov, and M. Belyaev are
with the Skolkovo Institute of Science and Technology, Moscow, 143026, Russia (e-mail: boris.shirokikh@skoltech.ru; alexey.shevtsov@skoltech.ru; m.belyaev@skoltech.ru). }

\thanks{A. Dalechina and V. Kostjuchenko are 
with the Moscow Gamma-Knife Center, Moscow, 125047, Russia (e-mail: adalechina@nsi.ru; vkostjuchenko@nsi.ru).}

\thanks{E. Krivov is 
with the Moscow Institute of Physics and Technology, Dolgoprudny, Moscow Region, 141701, Russia (e-mail: e.krivov@frtk.ru).}

\thanks{A. Durgaryan, M. Galkin, and A. Golanov are with 
the N.N. Burdenko National Medical Research Center of Neurosurgery, Moscow, 125047, Russia (e-mail: adurgaryan@nsi.ru; mgalkin@nsi.ru; golanov@nsi.ru).}

\thanks{\textit{This work has been submitted to the IEEE for possible publication. Copyright may be transferred without notice, after which this version may no longer be accessible}.}}

\maketitle

\begin{abstract}
We systematically evaluate a Deep Learning (DL) method in a 3D medical image segmentation task. Our segmentation method is integrated into the radiosurgery treatment process and directly impacts the clinical workflow. With our method, we address the relative drawbacks of manual segmentation: high inter-rater contouring variability and high time consumption of the contouring process. The main extension over the existing evaluations is the careful and detailed analysis that could be further generalized on other medical image segmentation tasks. Firstly, we analyze the changes in the inter-rater detection agreement. We show that the segmentation model reduces the ratio of detection disagreements from $0.162$ to $0.085$ $(p < 0.05)$. Secondly, we show that the model improves the inter-rater contouring agreement from $0.845$ to $0.871$ surface Dice Score $(p < 0.05)$. Thirdly, we show that the model accelerates the delineation process in between $1.6$ and $2.0$ times $(p < 0.05)$. Finally, we design the setup of the clinical experiment to either exclude or estimate the evaluation biases, thus preserve the significance of the results. Besides the clinical evaluation, we also summarize the intuitions and practical ideas for building an efficient DL-based model for 3D medical image segmentation.
\end{abstract}


\section{Introduction}
\label{sec:intro}

Advances in Deep Learning (DL) provide a solid foundation for automating medical image segmentation tasks \cite{lee2017deep}. Segmentation algorithms are successfully used to compute clinically relevant anatomical or disease characteristics. Furthermore, these algorithms, Convolutional Neural Networks (CNN) in particular, achieve near human-level performance in various tasks: a brain tumor, lung cancer, organ, organ-at-risk, or liver tumor segmentation; and actively continue to develop. The challenges and benchmark datasets, e.g., Brain Tumor Segmentation challenge \cite{bakas2018identifying}, mainly accelerate this development where the algorithms improve over each other in terms of the computer science metrics. More specifically, the standard way to evaluate the method is to use the metrics that only measure the similarity between the algorithm predictions and the ground truth annotations. However, the state-of-the-art performance on the benchmark data does not necessarily indicate a clinical effect. 

Although the development of such methods is a well-studied area, little research evaluates the effect of DL on clinical practice \cite{liu2019comparison}. Our work aims to close this gap and study how a carefully developed DL-based segmentation method contributes to the clinical workflow. The majority of these segmentation methods are applied to the radiological imaging and radiation therapy tasks \cite{sahiner2019deep}. In radiation therapy, a high dose of radiation should be precisely delivered to the considered targets. Due to the steep dose fall-off outside the target, the efficacy and safety of the treatment are highly dependent on the tumor, organ-at-risk (OAR), and critical structures contouring. Therefore, the required precision of radiation therapy provides a relevant background for a systematic evaluation of the segmentation method. 

We validate our method as the tool that helps to solve segmentation task in a clinically applicable way. We apply the developed segmentation model to stereotactic radiosurgery (SRS) treatment planning of multiple brain metastases. SRS is a single fraction radiation treatment in which a high dose of radiation is delivered to a precisely located target. SRS is an effective treatment option for patients with different brain disorders, including multiple brain metastases. However, manual delineation is a subjective process, and there is still no consensus in target volume delineation. In terms of subjectivity, tumor contouring is one of the weakest links in stereotactic radiosurgery \cite{njeh2008tumor,torrens2014standardization}. Furthermore, slice-by-slice manual segmentation could be very time-consuming, especially in the case of multiple lesions. In our work, we systematically evaluate how a DL-based segmentation method addresses all of these problems by improving contouring agreement and reducing contouring time.

Below, we discuss the most relevant works on automatic segmentation and clinical evaluation of DL methods, then summarize our contributions.

\subsection{Related work}


Dozens of recent studies have developed DL-based approaches for brain lesion segmentation and brain metastases segmentation in particular. To increase the segmentation quality, authors of \cite{liu2017deep,gonella2019investigating,bousabarah2020deep,grovik2020deep,shirokikh2021accelerating} suggested to design improved convolutional neural network (CNN) architectures. For the same reason, authors of \cite{charron2018automatic} suggested using a combination of different MRI modalities. In \cite{krivov2018tumor}, the improved patch sampling technique was proposed. Finally, several works \cite{shirokikh2019deep,hu2019multimodal,shirokikh2020universal} proposed to reweight a loss function to achieve higher segmentation scores.

Besides the segmentation score improvement, authors of \cite{lustberg2018clinical} compared the network predictions with the user adjustment of these generated contours as well as measured the time savings from the DL-assisted contouring. The authors used the contours from a single expert, thus intentionally did not study the inter-rater variability. Further, authors of \cite{schreier2020clinical} evaluated a DL method for male pelvis segmentation. The segmentation from clinical practice and the network predictions were assigned with the quality scores in their approach. In \cite{wong2020comparing}, authors compared DL-based segmentation of OAR to the expert inter-observer variability with Dice Score and Hausdorff distance. Summarizing the latter evaluations, the authors considered only the quality of the DL method's outputs. Instead of analysing the raw outputs, we evaluate their assistance effect on the clinical workflow.

Authors of \cite{zabel2021clinical} compared times needed to adjust manual, DL-generated, and atlas-based contours in OAR delineation for prostate radiation therapy. In addition to the time comparison, we extend our study with the evaluation of the inter-rater contouring variability. Both characteristics were evaluated in \cite{chlebus2019reducing} for the liver segmentation task and in \cite{bi2019deep} for the lung cancer segmentation task. However, the authors of the latter studies compared contours to the referenced ground truth generated by the other radiation oncologists. Contrary to their approaches, we evaluate the contouring variability for every rater independently, excluding possible biases from the reference to the side-expert ground truth. 

With this paper, we also extend our previous work on clinical evaluation of a DL model for the brain metastases segmentation \cite{shirokikh2019deep}. Firstly, we add the second study group with the reversed order of contouring techniques (manual and assisted) to evaluate the bias from the rater delineating the same image twice. Secondly, we study the detection agreement between raters. More recently, authors of \cite{lu2021randomized} confirmed the message of \cite{shirokikh2019deep} for non-expert radiologists and extended it with the detection analysis. However, the authors also use side-experts ground truth, which biases from the inter-rater contouring and detection variability measures. Contrary to the previous works, we also explicitly study the contouring agreement with the surface and volume metrics with the emphasis on the surface metric which is more sensitive to contouring. Moreover, with the careful experimental design, we additionally estimate and discuss detection, contouring, and time evaluation biases.

\subsection{Contributions}

Our work aims to assess the impact of the automatically generated contours by a convolutional neural network (CNN) on a clinical workflow. We show an improvement in inter-rater consistency and time savings from the usage of CNN-generated contours on a separate dataset with $20$ cases of brain metastatic lesions. With our methodology, we extend previous works and complement them in a novel way. We summarize our four main contributions in the clinical evaluation as follows:

\begin{itemize}
    \item Firstly, we evaluate changes in the inter-rater detection agreement. We show that CNN is likely to reduce the detection errors ratio by two times, from $0.162$ to $0.085$ ($p < 0.05$). Details could be found in Sec. \ref{sec:detection}.
    
    \item Secondly, we show the statistically significant improvement in the inter-rater contouring agreement from $0.845$ to $0.871$ surface Dice Score ($p < 0.05$) when our CNN model assists raters. The relevance of metrics, as well as the evaluation setup and results, are discussed in Sec. \ref{sec:contouring}.
    
    \item Thirdly, we show the statistically significant time savings. The assistance of our CNN model accelerates the delineation process up to two times on average and saves up to $5$ minutes of absolute delineation time on average ($p < 0.05$). These results, evaluation methodology, and the correspondence with contouring agreement are discussed in Sec. \ref{sec:time}.
    
    \item Finally, we use a carefully designed experimental setup dividing our dataset into two parts and evaluating manual and CNN-assisted contouring in a different order, see details in Sec. \ref{ssec:materials:contouring}. The latter allows us to estimate the biases of the same image being delineated twice, and preserve the significance of the results.
\end{itemize}






\section{Materials and Methods}
\label{sec:method}

\subsection{Datasets}
\label{ssec:methods:datasets}

To develop, train, and validate a CNN architecture, we used contrast-enhanced T1-weighted MR images of the patients treated in the Moscow Gamma Knife Center at the  N.N. Burdenko National Medical Research Center of Neurosurgery. The MR images were acquired on the $1.5$ Tesla MRI scanner (GE Healthcare). The image resolution was $0.9375 \times 0.9375 \times 1$ mm. We restored the original images and contours from the DICOM format from the archive of the Leksell Gamma Plan (LGP) database (Elekta).

\subsubsection{Development dataset}

Development dataset for training and validation consists of patients treated with Gamma Knife radiosurgery between $2005$ and $2011$. Some of these patients had several radiosurgical procedures. We include in the development dataset only the first procedure for each patient. We also include only the most common diagnoses (covering $77\%$ of cases): multiple metastases, meningioma, and acoustic schwannoma. Moreover, we exclude all patients with extracerebral metastases (e.g., skull lesions, orbital and optic nerve metastases). The lesions that not typically or low enhanced on T1-weighted were also excluded from the dataset. The reason to ignore such metastatic lesions is to develop a robust algorithm for the most typical cases. 


The final version of the training-validation part of the development dataset consists of $1420$ unique patients: $778$ cases of metastatic lesions, $371$ cases of meningioma, and $271$ cases of acoustic schwannoma. We use this data in a $3$-fold cross-validation setup to validate different versions of our CNN model. We additionally select $84$ unique patients who underwent radiosurgery in $2017$ for the final testing of our CNN model.


\subsubsection{Clinical dataset}

We perform a clinical evaluation of the automatic segmentation algorithm on a separate dataset. It consists of $20$ patients with multiple metastases who underwent radiosurgery from $2018$ to $2019$. The characteristics of selected patients are given in Tab. \ref{tab:clinical_data}. The number of lesions in Tab. \ref{tab:clinical_data} is strictly defined by the number of targets treated with radiosurgery. Note that these numbers may differ from the numbers obtained from contouring records.

\begin{table}[h!]
    \centering
    \caption{Patients' characteristics from the \textit{clinical dataset}. Group $1$: the starting point of delineation is assisted contouring (AC). Group $2$: the starting point of delineation is manual contouring (MC). The median number of lesions is $5$ in each group. The median total volume of lesions is $3.93$ cc in Group $1$ and $3.98$ cc in Group $2$.} 
    \begin{tabular}{c c c c c}
        \multirow{2}{*}{Case} & \multirow{2}{*}{Gender} & Histology of & Number of & \multirow{2}{*}{Total volume (cc)} \\
         & & primary tumor & lesions  &  \\
        \midrule
        Group 1 \\
        \bottomrule
        $\#1$ & female & stomach & $9$ & $1.56$ \\
        $\#2$ & male & renal & $7$ & $7.95$ \\
        $\#3$ & female & lung & $4$ & $3.90$ \\
        $\#4$ & female & lung & $3$ & $4.08$ \\
        $\#5$ & female & breast & $3$ & $0.57$ \\
        $\#6$ & male & lung & $10$ & $3.95$ \\
        $\#7$ & male & lung & $4$ & $0.45$ \\
        $\#8$ & male & renal & $7$ & $1.26$ \\
        $\#9$ & female & melanoma & $6$ & $8.69$ \\
        $\#10$ & male & renal & $4$ & $4.74$ \\
        \midrule
        Group 2 \\
        \bottomrule
        $\#1$ & male & lung & $11$ & $2.10$ \\
        $\#2$ & female & lung & $7$ & $7.42$ \\
        $\#3$ & female & breast & $4$ & $4.16$ \\
        $\#4$ & female & cervical & $4$ & $4.17$ \\
        $\#5$ & male & lung & $4$ & $0.47$ \\
        $\#6$ & female & lung & $11$ & $3.80$ \\
        $\#7$ & male & lung & $5$ & $0.50$ \\
        $\#8$ & male & melanoma & $5$ & $1.45$ \\
        $\#9$ & male & stomach & $8$ & $8.80$ \\
        $\#10$ & male & lung & $4$ & $4.83$ \\
       
    \end{tabular}
    \label{tab:clinical_data}
\end{table}

We divide these patients into two groups stratified by the median number and median volume of the lesions. These two groups correspond to different scenarios of tumor delineation, which are detailed below in Sec. \ref{ssec:materials:contouring}.

\subsection{Contouring Techniques}
\label{ssec:materials:contouring}

Four specialists from the Moscow Gamma Knife Center and Department of Stereotactic Radiation therapy at the N.N. Burdenko National Medical Research Center of Neurosurgery participated in this study: a senior medical physicist with $15$ years of experience in Gamma Knife radiosurgery (\textit{Rater 1}), a radiation oncologist with $7$ years of experience in Gamma Knife radiosurgery (\textit{Rater 2}), a radiation oncologist with $4$ years of experience in stereotactic radiation therapy (\textit{Rater 3}), and a radiation oncologist with $12$ years of experience in stereotactic radiation therapy (\textit{Rater 4}). All experts are actively involved in the routine treatment planning of brain tumors.

The experts delineated $20$ cases with multiple metastases divided into two groups. In the first trial in Group $1$, the experts used CNN-generated contours as a starting point of the delineation process and then adjusted these contours manually (if needed). We call this delineation technique \textit{assisted contouring (\textbf{AC})}. The delineation experiment was repeated after one week. In the second trial in Group $1$, the experts delineated tumors manually. We call this delineation technique \textit{manual contouring (\textbf{MC})}. In Group $2$, we conducted the same delineation experiments but \textit{changed the order of trials}: the first trial was \textit{MC} and the second trial one week later was \textit{AC}. The experts performed the delineation experiments independently and were blinded for the delineation results of each other in both groups and using both techniques (\textit{AC} and \textit{MC}).

In our previous paper \cite{shirokikh2019deep}, we have evaluated metric differences only within Group 2 setup (\textit{MC} first, \textit{AC} second). However, the question is whether the inter-rater agreement improvement and the time savings \textit{are caused by using the CNN-generated contours in the radiosurgery workflow} or whether the observed improvements \textit{are associated with the delineation repeated twice on the same MRI scans}? With this study we aim to resolve the uncertainty of double delineation by comparing two setups: Group $1$ (\textit{AC} first, \textit{MC} second) and Group $2$ (\textit{MC} first, \textit{AC} second).
 
All contouring experiments were performed in Leksell Gamma Plan (version $11.1$, Elekta AB) by Rater $1$ and Rater $2$ and in iPlan (version $4.5$, BrainLab) by Rater $3$ and Rater $4$. The DICOM data and CNN-generated radiation therapy (RT) structures were imported into the treatment planning systems.
Manual delineation was performed using standard tools for tumor segmentation enabled in the treatment planning systems.

For each case (one patient) in both groups, we recorded times required for both the adjustment of the CNN-initialized contours (\textit{AC}) and manual contouring (\textit{MC}).

\subsection{Deep Learning Model}
\label{ssec:methods:cnn}

Recently, convolutional neural networks (CNN) have become the dominant approach to solve medical image segmentation tasks \cite{meyer2018survey}. Therefore, we further focus on solving our binary segmentation problem only via training a CNN.

\begin{figure*}[h]
    \centering
    \includegraphics[width=\linewidth]{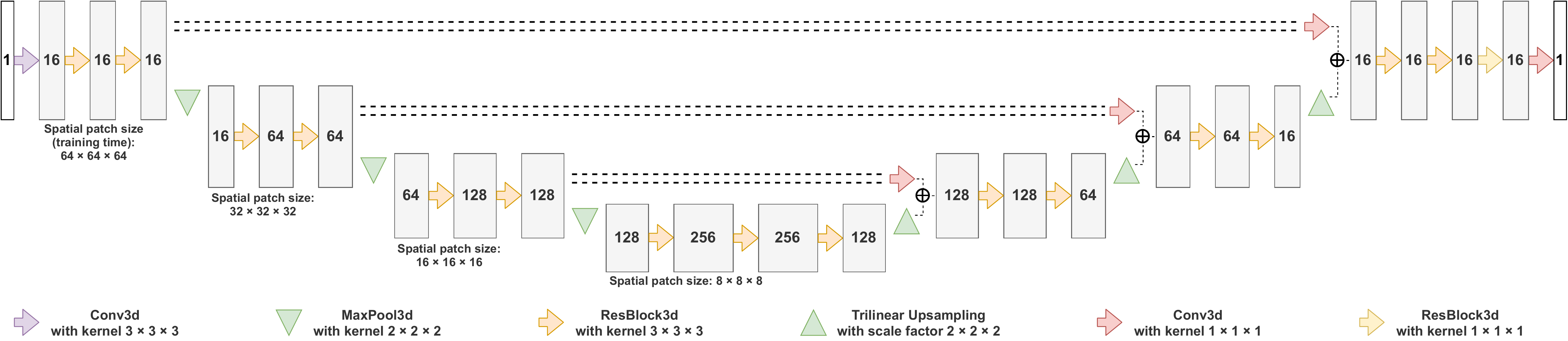}
    \caption{The architecture of the CNN model integrated into the clinical workflow.}
    \label{fig:cnn}
\end{figure*}

Practical application with its limitations on memory consumption and segmentation quality force us to develop a CNN model with several modifications and modify the training procedure. We start with building the CNN architecture (see Fig. \ref{fig:cnn}) that uses the most relevant state-of-the-art techniques. Our architecture is de facto $3$D U-Net \cite{cciccek20163d} that has been proven to be one of the best models for $2$D \cite{ronneberger2015u} and $3$D \cite{milletari2016v} medical image segmentation tasks. During the cross-validation experiments, we obtain better results by using residual blocks \cite{he2016deep}. Also, we change the channel-wise concatenation at the end of the shortcuts to the channel-wise summation and apply convolutional layers with $1 \times 1 \times 1$ kernels to all shortcuts. Finally, we limit the number of filters in the original resolution to $16$ to satisfy the memory constraints of the clinical software.

We find that the further improvement of the architecture does not benefit the quality. This observation also highly correlates with the suggestion of \cite{isensee2018no} to focus on building the appropriate training procedure. Therefore, we further address task-specific circumstances, such as the relatively small size of multiple metastases and differences in tumor diameters. Firstly, in \cite{krivov2018tumor} we introduced a non-uniform patch sampling technique for CNN training called Tumor Sampling. Instead of sampling from the foreground class as in \cite{kamnitsas2017efficient}, we equally represent every tumor by uniform sampling from the tumor instances. Thus, Tumor Sampling considerably increases the object-wise Recall of the model, especially for the small tumors, which is more carefully evaluated in \cite{shirokikh2019deep}. Secondly, we supplement the Binary Cross-Entropy (BCE) loss function with the inverse weighting strategy \cite{shirokikh2020universal}. Inverse weighting assigns larger weights to smaller lesions (inversely proportional to the lesion volume), thus further increasing the CNN's ability to detect small lesions.

During the training and inference, we scale the input MR images to have intensities between $0$ and $1$. We do not apply any other preprocessing steps. The final version of the integrated model is trained for $100$ epochs, starting with the learning rate of $10^{-2}$, and reducing it to $10^{-3}$ at the epoch $80$. Each epoch consists of $100$ iterations of stochastic gradient descent with Nesterov momentum ($0.9$) minimizing inversely weighted Binary Cross-Entropy. At every iteration, we sample patches of size $64 \times 64 \times 64$ and batch size of $16$ via the Tumor Sampling strategy. We also find the fixed training policy enough for the optimizer to converge, and adjusting the training policy does not benefit the cross-validation score.

The training takes about $8$ hours on a $24$GB NVIDIA Tesla M$40$ GPU. The inference takes place on a $6$GB NVIDIA GeForce GTX $1060$ within the clinical workflow, which processes an entire image in a one forward step. The inference time is typically less than $5$ seconds.

\section{Improving the Inter-rater Detection Agreement}
\label{sec:detection}

The analysis of delineation agreement is available only when exactly $4$ out of $4$ raters detect a tumor. However, the problem is that some regions of interest (ROI) are delineated only by $3$, $2$, or even $1$ raters. Therefore, we start by assessing the impact of CNN on the inter-rater detection agreement.

Firstly, we define the ground truth (GT) tumors of three types:
\begin{itemize}
    \item $\mathbf{GT}$ -- tumors that are delineated with \textit{MC} by at least $3$ raters.
    \item $\mathbf{GT^+}$ -- tumors that are delineated with \textit{AC} by at least $3$ raters.
    \item $\mathbf{GT^*}$ -- tumors that are included either in $GT$ or in $GT^+$. $GT^* = GT \cup GT^+$.
\end{itemize}

Further, we refer to $\mathbf{GT^*}$ as a golden standard for tumor detection for both \textit{MC} and \textit{AC}. With the golden standard, we could define every delineated tumor as True Positive (TP) or False Positive (FP) and define the missed $\mathbf{GT^*}$ tumor as False Negative (FN).

\begin{itemize}
    \item $\mathbf{TP}$ -- \textit{MC} tumor which has non-zero intersection with $\mathbf{GT^*}$ tumors.
    \item $\mathbf{FP}$ -- \textit{MC} tumor which has no intersection with $\mathbf{GT^*}$ tumors.
    \item $\mathbf{FN}$ -- $\mathbf{GT^*}$ tumor which has no intersection with \textit{MC} tumors.
\end{itemize}

Then $\mathbf{TP^+}$, $\mathbf{FP^+}$ and $\mathbf{FN^+}$ are defined the same way for every single rater using \textit{AC}. The $\mathbf{GT^*}$ tumors also remain the golden standard for the latter case. Note that True Negative (TN) detections are undefined in our task.

Finally, we could compare \textit{MC} and \textit{AC} techniques by the number of valid detections ($\mathbf{TP}$ vs. $\mathbf{TP^+}$) and the number of errors ($\mathbf{FP} + \mathbf{FN}$ vs. $\mathbf{FP^+} + \mathbf{FN^+}$). In most cases, a rater manually delineates the image in two steps: (i) reviews each slice of the MRI image, then (ii) examines every suspected ROI, whether it is a lesion or not. Therefore, we assume that all TP, FP, and FN outcomes have approximately the same nature: in the first stage, a rater saw all ROIs and then decide on the contouring of the lesion. Assuming the same nature of TP, FP and FN, we model the contouring process as Bernoulli random variables: $0$ is the TP (correct) outcome, $1$ is FP or FN (incorrect) outcome. Thus, the probability to incorrectly delineate a tumor is the parameter of Bernoulli distribution: $\mathbf{p_{err}}$ for \textit{MC} technique and $\mathbf{p_{err}^+}$ for \textit{AC} technique. We use one-sided Z-test as an approximation to Student's T-test (we have $> 50$ samples in each case) to test whether raters make fewer detection mistakes using \textit{AC} or not. It has the following null hypothesis:

\begin{equation}\label{eq:hyp_bernoulli}
    \begin{array}{l}
        H_0: p_{err}^+ = p_{err} \text{ \textit{vs.}} \\
        H_1: p_{err}^+ < p_{err}.
    \end{array}
\end{equation}

Here, the estimators for these parameters are $\hat{p}_{err} = N_{err} / (TP + N_{err})$ and $\hat{p}_{err}^+ = N_{err}^+ / (TP^+ + N_{err}^+)$. And the numbers of errors are defined as $N_{err} = FN + FP$ and $N_{err}^+ = FN^+ + FP^+$.

\subsection{Merging groups in detection agreement evaluation}
\label{ssec:detection:merging}

We use averaged raters' opinion to perform a more robust comparison between \textit{MC} and \textit{AC} techniques. The latter means that we test the hypothesis (Eq. \ref{eq:hyp_bernoulli}) of Z-test by averaging TP, FP, and FN values of $4$ raters. 
Tab. \ref{tab:detection_global} shows averaged TP, FP, and FN values as well as calculated estimator's values and P-value in $3$ setups: within Group $1$, within Group $2$ and merging both groups. Note that calculation of z-score and estimators in Z-test could be easily extended to non-integer values of correct or incorrect outcomes without any restrictions. We use this generalization of Z-test along with the assumption that the ``average rater'' follows the same Bernoulli random process.

\begin{table*}[h!]
    \centering
    \caption{Detection agreement for the average raters' opinion.}
    \begin{tabular}{l c c c c c c c c c c c}
        \toprule
        & \multicolumn{5}{c}{Manual contouring (\textit{MC})} & \multicolumn{5}{c}{Assisted contouring (\textit{AC})} & \multirow{2}{*}{P-value ($\hat{p}_{err}$ \textit{vs.} $\hat{p}_{err}^+$)} \\
        \cmidrule(lr){2-6}
        \cmidrule(lr){7-11}
         & $FN$ & $FP$ & $N_{err}$ & $TP$ & $\hat{p}_{err}$ & $FN^+$ & $FP^+$ & $N_{err}^+$ & $TP^+$ & $\hat{p}_{err}^+$ \\
        \midrule
        Group 1  & $4.00$ & $4.25$ & $8.25$ & $54.00$ & $0.133$ & $3.50$ & $2.50$ & $6.00$ & $54.50$ & $0.099$ & $0.287$ \\
        \midrule
        Group 2  & $5.00$ & $7.25$ & $12.25$ & $52.00$ & $0.191$ & $0.75$ & $3.50$ & $4.25$ & $56.25$ & $0.070$ & $0.026$ \\
        \midrule
        Merged & $9.00$ & $11.50$ & $20.50$ & $106.00$ & $0.162$ & $4.25$ & $6.00$ & $10.25$ & $110.75$ & $0.085$ & $0.036$ \\
        \bottomrule
    \end{tabular}
    \label{tab:detection_global}
\end{table*}

Both groups result in reduction of the detection error rate: $\Delta \hat{p}_{err} = 0.034$ for Group $1$ and $\Delta \hat{p}_{err} = 0.121$ for Group $2$, where $\Delta \hat{p}_{err} = \hat{p}_{err} - \hat{p}_{err}^+$. Despite we have a positive $\Delta \hat{p}_{err}$, the magnitude of this result heavily depends on the techniques order within the groups. Group $2$ with the larger delta has \textit{MC} as a starting point and then it repeats delineation using \textit{AC}. Raters have seen the same image twice which could possibly create bias towards the lower error rate in the second trial of contouring. Therefore, in Group $2$ it could underestimate $\hat{p}_{err}^+$, thus overestimate $\Delta \hat{p}_{err}$. Similarly, in Group $1$ it could underestimate $\hat{p}_{err}$, thus underestimate $\Delta \hat{p}_{err}$. This pattern could be seen better with the calculated values: $\Delta \hat{p}_{err} = 0.034$ for Group $1$, where we assume the underestimation of $\Delta \hat{p}_{err}$, and $\Delta \hat{p}_{err} = 0.121$ for Group $2$, where we assume the overestimation of $\Delta \hat{p}_{err}$. However, there is not enough evidence at $5\%$ significance level to state that we have significant time reduction within Group $1$.

We remove the uncertainty of double delineation by testing the same hypothesis (Eq. \ref{eq:hyp_bernoulli}) on merged data from both groups, see the last row of Tab. \ref{tab:detection_global}. Therefore, we have enough evidence at $5\%$ significance level to state, that CNN reduces the probability of incorrect detection from $0.162$ to $0.085$ ($\Delta \hat{p}_{err} = 0.077$).

\subsection{Detailing detection agreement for every rater}
\label{ssec:detection:detailed}

In Tab. \ref{tab:detection_raters} we give more detailed results for every rater and for the original CNN predictions. Moreover, we supplement reported TP, FP and FN with more interpretable metrics: \textit{Recall} and \textit{average FP} (avg FP). $Recall = \frac{TP}{TP + FN} = \frac{TP}{ |GT^*|}$, thus Recall gives us the fraction of tumors that a specific rater or CNN detected. Average FP is the average number of False Positives per image that a specific rater or CNN misdetected.

Since the hypothesis is tested for every rater (repeated $4$ times) within the same framework, we apply multiple testing correction (MTC). We use Benjamini-Hochberg correction with the significance level $\alpha = 0.05$. Instead of iteratively correcting the significance level and comparing it with the sorted P-values, we correct the sorted P-values on the constant significance level. It is an equivalent representation of the multiple testing correction. The corrected P-value is $\tilde{p}_{(i)} = m p_{(i)} / i$, where $p_{(1)} \leq \dots \leq p_{(m)}$ are the sorted P-values and $m=4$ is the number of tested hypotheses.

However, corrected P-values give us enough evidence at $5\%$ significance level to state that CNN reduces the probability of false detection only for Rater $2$. Despite $\Delta \hat{p}_{err}$ is positive for all raters, the size of the evaluation dataset should be increased to obtain statistically significant results for every selected rater.

\begin{table*}[h]
    \centering
    \caption{Detection agreement for every rater and CNN. The last column presents the corrected P-values via Benjamini-Hochberg multiple testing correction (MTC) at the significance level $\alpha = 0.05$. The order number $(i)$ of the sorted P-values is given in the brackets.}
    \begin{tabular}{l c c c c c c c c}
        \toprule
        & Recall & Recall$^+$ & avg FP & avg FP$^+$ & $\hat{p}_{err}$ & $\hat{p}_{err}^+$ & P-value ($\hat{p}_{err}$ \textit{vs.} $\hat{p}_{err}^+$) & MTC of P-value \\
        \toprule
        Rater 1 & $.939$ & $.974$ & $.600$ & $.400$ & $.150$ & $.089$ & $.079$ & $.105 \,\, (3)$ \\
        \midrule
        Rater 2 & $.896$ & $.948$ & $.700$ & $.150$ & $.202$ & $.076$ & $.003$ & $.010 \,\, (1)$ \\
        \midrule
        Rater 3 & $.939$ & $.957$ & $.600$ & $.250$ & $.150$ & $.083$ & $.058$ & $.115 \,\, (2)$ \\
        \midrule
        Rater 4 & $.913$ & $.974$ & $.400$ & $.400$ & $.146$ & $.089$ & $.090$ & $.090 \,\, (4)$ \\
        \midrule
        CNN & $.896$ & --- & $.550$ & --- & $.183$ & --- & --- & --- \\
        \bottomrule
    \end{tabular}
    \label{tab:detection_raters}
\end{table*}

\subsection{Visualizing detection agreement}


Finally, we compare sizes of correct (tumors that are delineated by all raters using both \textit{MC} and \textit{AC}) and incorrect (the other tumors) delineations. We show that the incorrect delineations are significantly smaller in size than the correct delineations. The difference in size is tested using one-sided Mann-Whitney rank test (U-test) on these two samples. It has the following null and alternative hypotheses:

\begin{equation}\label{eq:hyp_mw_greater}
    \begin{array}{l}
        H_0: \mathbb{P} (X > Y) = \mathbb{P} (X < Y) \text{ versus} \\
        H_1: \mathbb{P} (X > Y) > \mathbb{P} (X < Y),
    \end{array}
\end{equation}

\noindent
where $\mathbb{P} (X > Y)$ is the probability of an observation from population $X$ exceeding an observation from population $Y$. In our case, population $X$ is the correct prediction sizes, and population $Y$ is the incorrect prediction sizes. Under the size we understand the average diameter of the tumor. Firstly, we average the volumes of the corresponding raters' masks, and then we calculate the diameter of the ball with that volume.

\begin{figure}[h]
    \centering
    \includegraphics[width=\linewidth]{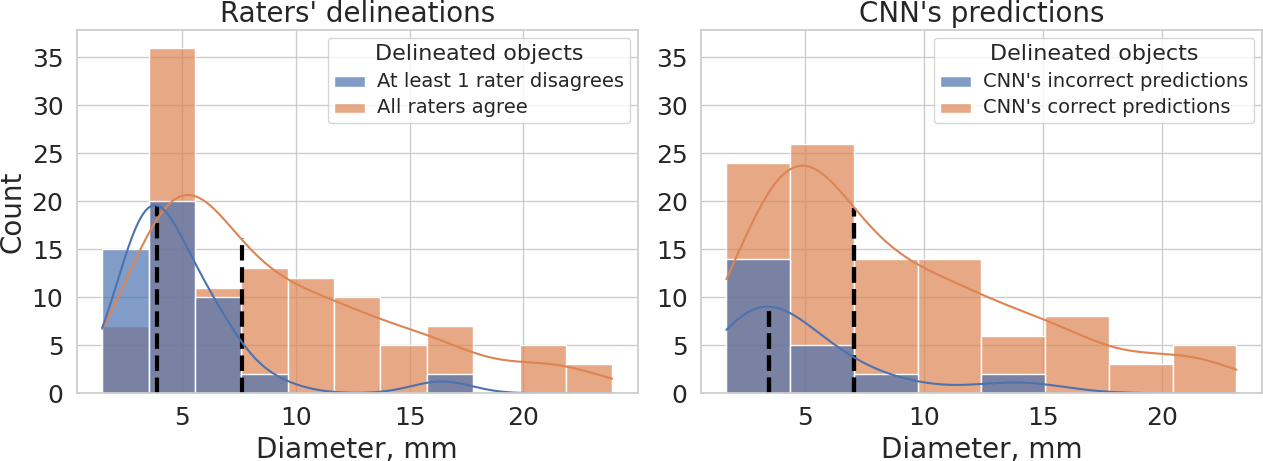}
    \caption{Diameters of the raters' delineations (left) and the CNN's predictions (right). Blue -- diameters of incorrect delineations and predictions; orange -- diameters of correct delineations and predictions. The black dashed lines are the median values of the corresponding diameter samples.}
    \label{fig:sizes}
\end{figure}

We further show that both the raters and CNN disagree on small tumors that might be considered as vessels or have a non-typical localization. Under incorrect delineations we assume such an example when at least one rater or CNN does not provide segmentation for the target. The histogram of diameters of the correctly and incorrectly detected targets is presented in Fig. \ref{fig:sizes} (left). The median diameter of incorrectly detected target is $3.9$ mm. The median diameter of correctly detected target is $7.6$ mm, which is almost two times larger. The P-value of U-test (Eq. \ref{eq:hyp_mw_greater}) is $9 \cdot 10^{-9}$, which gives us enough evidence at $5\%$ significance level to state that the incorrectly detected targets are smaller than the correctly detected targets. 

In Fig. \ref{fig:sizes} (right), we present the same analysis for CNN's predictions. However, in this case, we calculate the diameter of TPs and FPs using only CNN's masks. The diameter of FNs is calculated by averaging raters' masks, as earlier. Here, we have similar results: the median diameter of incorrect CNN's prediction is $3.5$ mm, and the median diameter of correct CNN's prediction is $7.0$ mm, which is two times larger. The corresponding P-value is $9 \cdot 10^{-5}$, which also gives us enough evidence at $5\%$ significance level to state that the incorrect CNN's predictions are smaller than the correct CNN's predictions.

\begin{figure*}[h]
    \centering
    \includegraphics[width=\linewidth]{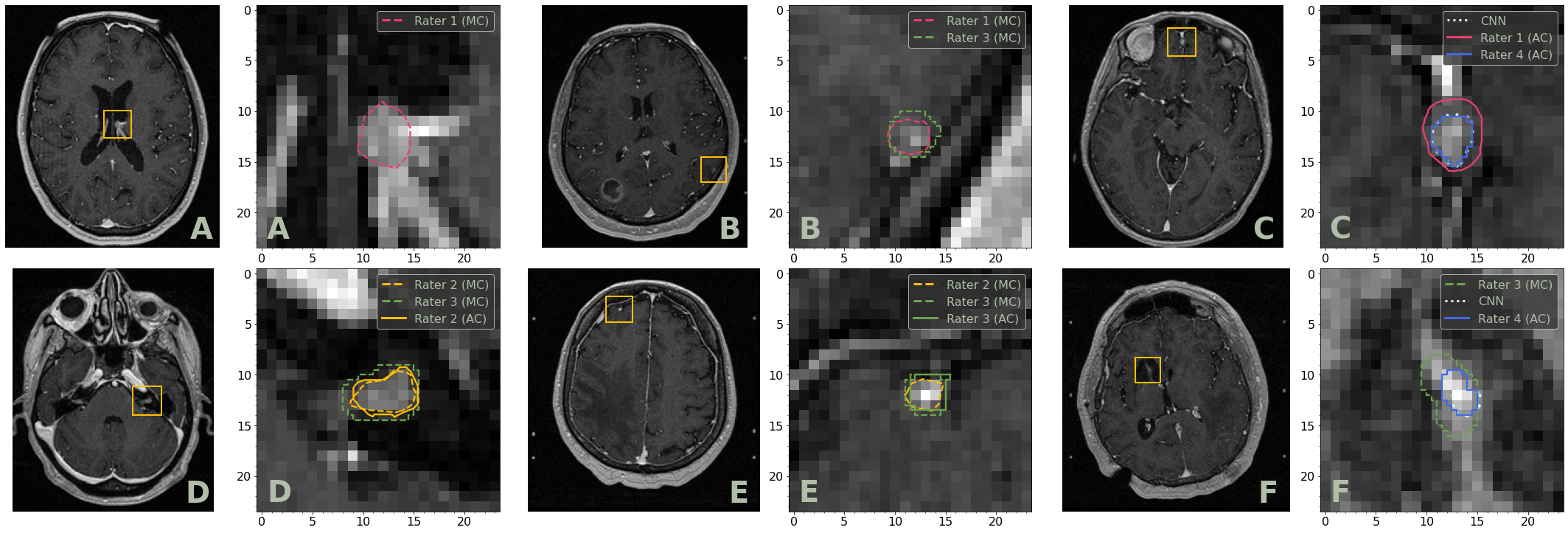}
    \caption{Examples of False Positive delineations made by raters or CNN. The falsely detected objects are mostly small tumors that might be considered as vessels or have a non-typical localization. In every pair, we show a 2D slice (left) and the zoomed region with the contours (right). In-plane MRI resolution is $0.9375 \times 0.9375$ mm. Only delineated contours are plotted.}
    \label{fig:fp}
\end{figure*}

\begin{figure*}[h]
    \centering
    \includegraphics[width=\textwidth]{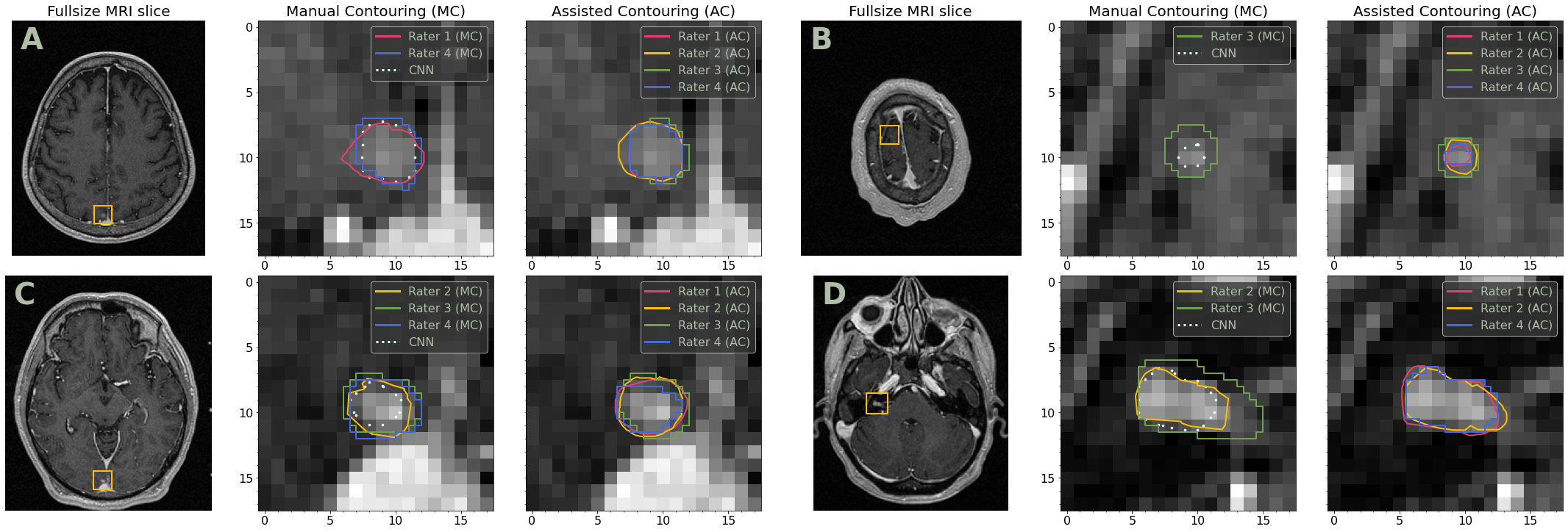}
    \caption{Examples of small tumors indicated by CNN that result in the False Negatives reduction. In every triplet, we show a 2D slice (left), the zoomed region with the \textit{manual contouring} and the initial CNN's contour (center), and the zoomed region with the \textit{assisted contouring} (right). In-plane MRI resolution is $0.9375 \times 0.9375$ mm. Only delineated contours are plotted.}
    \label{fig:fn}
\end{figure*}

Therefore, both CNN and raters tend to miss or mistakenly delineate small lesions. The examples of such False Positives are given in Fig. \ref{fig:fp}. The reduction of False Negatives (or the increase of Recall) when switching from \textit{MC} to \textit{AC} is shown in Fig. \ref{fig:fn}.

\section{Improving the Inter-rater Contouring Agreement}
\label{sec:contouring}

We further evaluate how CNN improves the contouring agreement. Contrary to the detection analysis, contouring analysis is available only when exactly $4$ out of $4$ raters delineate a tumor with both \textit{MC} and \textit{AC} techniques. Consequently, the number of tumors under analysis is reduced from $115$ to $83$ ($43$ in Group $1$ and $40$ in Group $2$).

To evaluate the \textit{MC} consistency for a particular rater, we compare the manual contour of that rater with the averaged manual contours of the other raters ($\mathbf{1}$ \textbf{vs.} $\mathbf{3}$). The \textit{AC} consistency is evaluated by comparing the adjusted contour with the same average manual contour of the three other raters ($\mathbf{1^+}$ \textbf{vs.} $\mathbf{3}$). In both ($\mathbf{1}$ \textbf{vs.} $\mathbf{3}$) and ($\mathbf{1^+}$ \textbf{vs.} $\mathbf{3}$) setups, we compare a single contour to the corresponding calculated reference via surface Dice Score \cite{nikolov2018deep} and concordance index.

Although most studies use Dice Score as a primary metric, we consider surface Dice Score our main metric. Surface Dice Score (sDSC) is designed to be sensitive to the surface difference \cite{nikolov2018deep}. Therefore it measures exactly the inter-rater agreement on the tumor boundaries, the main source of subjectivity. sDSC shares the same formula with Dice Score: $sDSC = 2 |A \cap B| / \left( |A| + |B| \right)$, where $|A|$ and $|B|$ are the surface areas of the given masks, and $|A \cap B|$ is the surface area of masks' intersection at the predefined tolerance distance. We use the tolerance distance equal to $1.0$ mm, as in \cite{vaassen2020evaluation}. Moreover, authors of \cite{vaassen2020evaluation} show a high correlation between sDSC improvement and the relative contouring time reduction.

On the other hand, we preserve the consistency with the previous works by using the concordance index (CCI) as a volumetric measure. $CCI = |AV_{3} \cap V_{obs}| / |AV_{3} \cup V_{obs}|$, where $V_{obs}$ is the tumor mask delineated by a specific rater and $AV_{3}$ is the average mask of three other raters. Given a specific of our task, Dice Score (DSC) could be written as $DSC = 2 |AV_{3} \cap V_{obs}| / \left( |AV_{3}| + |V_{obs}| \right)$, and thus it functionally depends on CCI, $DSC = 2 CCI / \left( 1 + CCI \right)$. Moreover, CCI is widely used to compare differently delineated target volumes in radiation therapy \cite{farace2011clinical,sandstrom2018multi}, therefore we further use CCI instead of DSC as a volumetric measure for contouring agreement.

Inside every group, we have quality metrics calculated in both \textit{MC} ($\mathbf{1}$ \textbf{vs.} $\mathbf{3}$) and \textit{AC} ($\mathbf{1^+}$ \textbf{vs.} $\mathbf{3}$) setups for every rater and for every tumor. Firstly, we test whether Group $1$ and Group $2$ could be merged or not. To do so, we calculate sDSC and CCI deltas between ($\mathbf{1^+}$ \textbf{vs.} $\mathbf{3}$) scores and ($\mathbf{1}$ \textbf{vs.} $\mathbf{3}$) scores. Then, we assume these deltas to be independently generated samples, thus they could be tested to have unequal distributions via TOST (two one-sided T-test). It has the following null and alternative hypotheses:

\begin{equation}\label{eq:hyp_tost}
    \begin{array}{l}
        H_0: \mu_U - \mu_L < \Delta_L \text{ or } \mu_U - \mu_L > \Delta_U \text { versus}  \\
        H_1: \Delta_L < \mu_U - \mu_L < \Delta_U ,
    \end{array}
\end{equation}

\noindent
where $\Delta_U$ and $\Delta_L$ are the upper and lower equivalence bounds, respectively. We set the bounds to the $\pm 1/4$ of the estimated standard deviation.

Therefore, we show that different order of \textit{MC} and \textit{AC} techniques in Group $1$ and Group $2$ does not affect the results of the inter-rater contouring agreement improvement. TOST test gives us $\text{P-value} = 0.011$ for sDSC deltas, and $\text{P-value} = 0.038$ for CCI deltas. Therefore, we have enough evidence at $5\%$ significance level to state that the deltas of sDSC and CCI follows the same distribution in Group $1$ and Group $2$.

Finally, we consider ($\mathbf{1^+}$ \textbf{vs.} $\mathbf{3}$) and ($\mathbf{1}$ \textbf{vs.} $\mathbf{3}$) scores as paired samples to test whether sDSC ($\mathbf{1^+}$ \textbf{vs.} $\mathbf{3}$) is higher than sDSC ($\mathbf{1}$ \textbf{vs.} $\mathbf{3}$) or not, and whether CCI ($\mathbf{1^+}$ \textbf{vs.} $\mathbf{3}$) is higher than CCI ($\mathbf{1}$ \textbf{vs.} $\mathbf{3}$) or not. We use one-sided Wilcoxon signed-rank test, that has the same hypotheses as Eq. \ref{eq:hyp_mw_greater}, but the population $X$ is the ($\mathbf{1^+}$ \textbf{vs.} $\mathbf{3}$) scores and the population $Y$ is the ($\mathbf{1}$ \textbf{vs.} $\mathbf{3}$) scores.

In the Tab. \ref{tab:contouring_q}, we report the increase of sDSC and CCI when the raters switch the technique from \textit{MC} to \textit{AC} along with the corresponding P-values. All raters pass the $5\%$ significance level, except Rater $1$. We discuss the possible reason for the contouring agreement of Rater $1$ not to be significantly improved in Sec. \ref{sec:discussion}.



\begin{table*}[h]
    \centering
    \caption{Inter-rater contouring agreement improvement in terms of median surface Dice Score (sDSC) and median concordance index (CCI). The ($1$ vs. $3$) and ($1^+$ vs. $3$) setups correspond to the \textit{MC} and \textit{AC} techniques, respectively. Paired sDSC and CCI samples are tested via one-sided Wilcoxon signed-rank test to show whether the metric value in the ($1$ vs. $3$) setup is lower than in the ($1^+$ vs. $3$) setup.}
    \begin{tabular}{l c c c c c c}
        \toprule
         & sDSC ($1$ vs. $3$) & sDCS ($1^+$ vs. $3$) & P-value$_{sDSC}$ & CCI ($1$ vs. $3$) & CCI ($1^+$ vs. $3$) & P-value$_{CCI}$ \\
        \midrule
        Rater 1  & .868 & .875 & 2.60 e-01 & .818 & .822 & 2.59 e-01 \\
        \midrule
        Rater 2  & .859 & .875 & 2.08 e-03 & .798 & .810 & 6.52 e-03 \\
        \midrule
        Rater 3  & .714 & .770 & 4.59 e-06 & .714 & .758 & 4.87 e-07 \\
        \midrule
        Rater 4  & .865 & .912 & 2.61 e-11 & .811 & .852 & 5.43 e-11 \\
        \midrule
        All data & .845 & .871 & 3.84 e-14 & .787 & .807 & 3.84 e-14 \\
        \bottomrule
    \end{tabular}
    \label{tab:contouring_q}
\end{table*}

\begin{figure*}[h!]
    \centering
    \includegraphics[width=\linewidth]{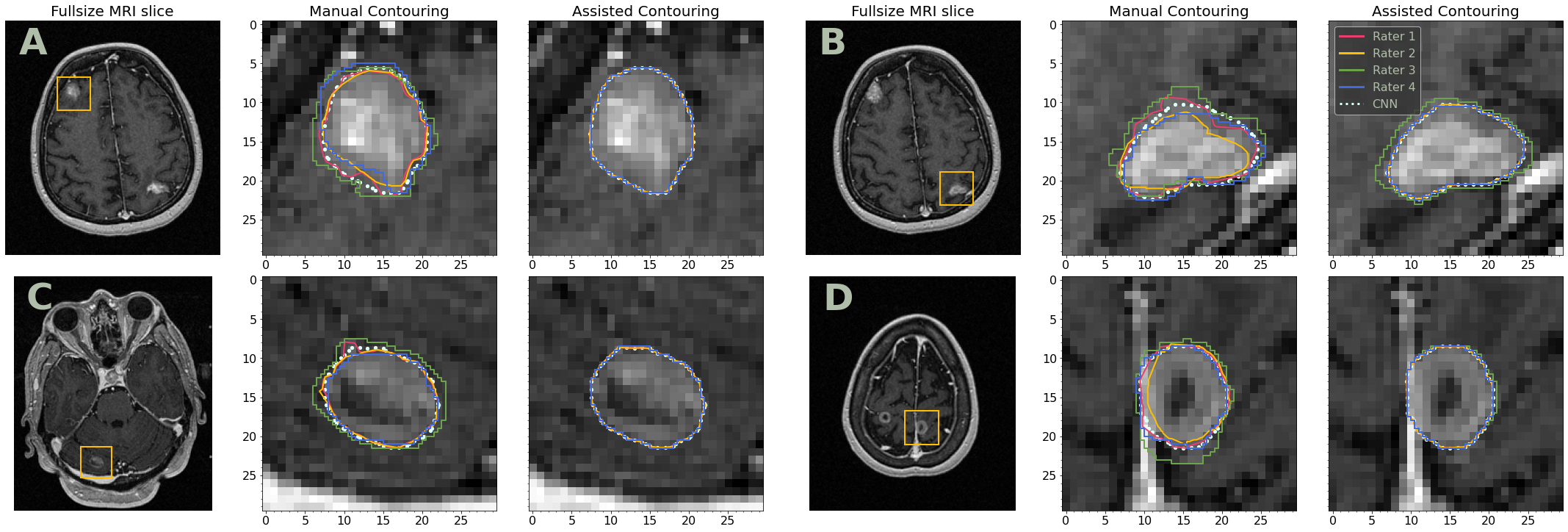}
    \caption{Examples of the metastatic lesions delineated by the raters manually (middle column) and with CNN assistance (right column). The white dotted contours correspond to the CNN's suggestions. In-plane MRI resolution is $0.9375 \times 0.9375$ mm.}
    \label{fig:contouring_agreement}
\end{figure*}

The last row of the Tab. \ref{tab:contouring_q} gives significant inter-rater contouring agreement improvement when the data for all raters is merged. Agreement is increased from $0.845$ (\textit{MC}) to $0.871$ (\textit{AC}) in terms of median sDSC, median $\Delta \text{sDSC} = 0.026$. Median CCI is increased from $0.787$ (\textit{MC}) to $0.807$ (\textit{AC}), median $\Delta \text{CCI} = 0.020$. Visual examples of contouring agreement improvement are given in Fig. \ref{fig:contouring_agreement}.

\section{Reducing the Raters' Contouring Time}
\label{sec:time}

We conclude our analysis by evaluating how CNN accelerates the contouring process. To show the benefit from the CNN usage, we calculate the relative time reduction when switching from \textit{MC} to \textit{AC} technique. Although the shown time reduction is positive, Group $1$ and Group $2$ have significantly different sample distributions, thus could not be merged. Below, we firstly test our groups to have different sample distributions and then show a significant time reduction inside each group.

We denote the time required for manual contouring (\textit{MC}) as $t_{mc}$ and the time required to adjust the CNN-initialized contours (\textit{AC}) as $t_{ac}$. Inside each group, we calculate relative time reduction $\Delta t_{rel} = \left( t_{mc} - t_{ac} \right) / t_{mc}$ for every case. Hence, we could test samples $\Delta t_{rel}$, whether they have the same distribution inside each group or not. Samples $\Delta t_{rel}$ from Group $1$ and Group $2$ are unpaired; therefore, we use one-sided Mann-Whitney rank test with the same hypotheses as in Eq. \ref{eq:hyp_mw_greater}. Here, population $X$ is $\Delta t_{rel}$ samples in Group $2$, and population $Y$ is $\Delta t_{rel}$ samples in Group $1$. With one-sided Mann-Whitney rank test, we get enough evidence (P-value $= 0.019$) at $5\%$ significance level to state that the two distributions are different. 

Although the distributions are different, we further demonstrate the time reduction independently within each group. To do so, we test the difference between $t_{mc}$ and $t_{ac}$ samples for each group. Since $t_{mc}$ and $t_{ac}$ are paired samples, we use one-sided Wilcoxon signed-rank test. It has the same hypothesis as Eq. \ref{eq:hyp_mw_greater}, where population $X$ is now $t_{mc}$ samples, and population $Y$ is $t_{ac}$ samples. In Tab. \ref{tab:contouring_t}, we show the median manual contouring time and the median contouring time reduction for every rater, and we also include the relative contouring speedup ($t_{mc} / t_{ac}$). Wilcoxon test's P-values give us enough evidence at $5\%$ significance level to state that the CNN-assisted contouring accelerates the delineation process for every rater.

By merging data for all raters (Tab. \ref{tab:contouring_t}, last row), we conclude that the CNN-assisted contouring speedups delineation process in between $1.6$ and $2.0$ times on average. The absolute median time reduction with the CNN-assisted contouring is between $03$:$26$ and $04$:$53$ (\textit{mm:ss}).


\begin{table*}[h]
    \centering
    \caption{The median time of \textit{manual contouring} ($t_{mc}$), median contouring time reduction via \textit{assisted contouring} ($t_{mc} - t_{ac}$), and median relative speedup via \textit{assisted contouring} ($t_{mc} / t_{ac}$). Absolute times are given in \textit{mm:ss} format. We also test whether the $t_{ac}$ samples are significantly smaller than the $t_{mc}$ samples via one-sided Wilcoxon signed-rank test and report P-values.}
    \begin{tabular}{l c c c c c c c c}
        \toprule
         & \multicolumn{4}{c}{Group 1} & \multicolumn{4}{c}{Group 2} \\
        \cmidrule(lr){2-5}
        \cmidrule(lr){6-9}
         & $t_{mc}$ & $t_{mc} - t_{ac}$ & $t_{mc} / t_{ac}$ & P-value & $t_{mc}$ & $t_{mc} - t_{ac}$ & $t_{mc} / t_{ac}$ & P-value \\
        \midrule
        Rater 1  & 09:58 & 04:41 & 2.09 & 2.53 e-03 & 10:35 & 06:45 & 2.85 & 2.53 e-03 \\
        \midrule
        Rater 2  & 05:46 & 01:28 & 1.48 & 1.61 e-02 & 05:20 & 01:37 & 1.67 & 2.53 e-03 \\
        \midrule
        Rater 3  & 09:38 & 05:03 & 2.19 & 3.46 e-03 & 13:12 & 06:24 & 2.13 & 2.53 e-03 \\
        \midrule
        Rater 4  & 07:39 & 02:00 & 1.36 & 2.53 e-03 & 11:43 & 05:44 & 1.91 & 2.53 e-03 \\
        \midrule
        All data & 08:38 & 03:26 & 1.60 & 8.84 e-08 & 10:17 & 04:53 & 2.01 & 1.78 e-08 \\
        \bottomrule
    \end{tabular}
    \label{tab:contouring_t}
\end{table*}

\begin{figure*}[h!]
    \centering
    \includegraphics[width=\linewidth]{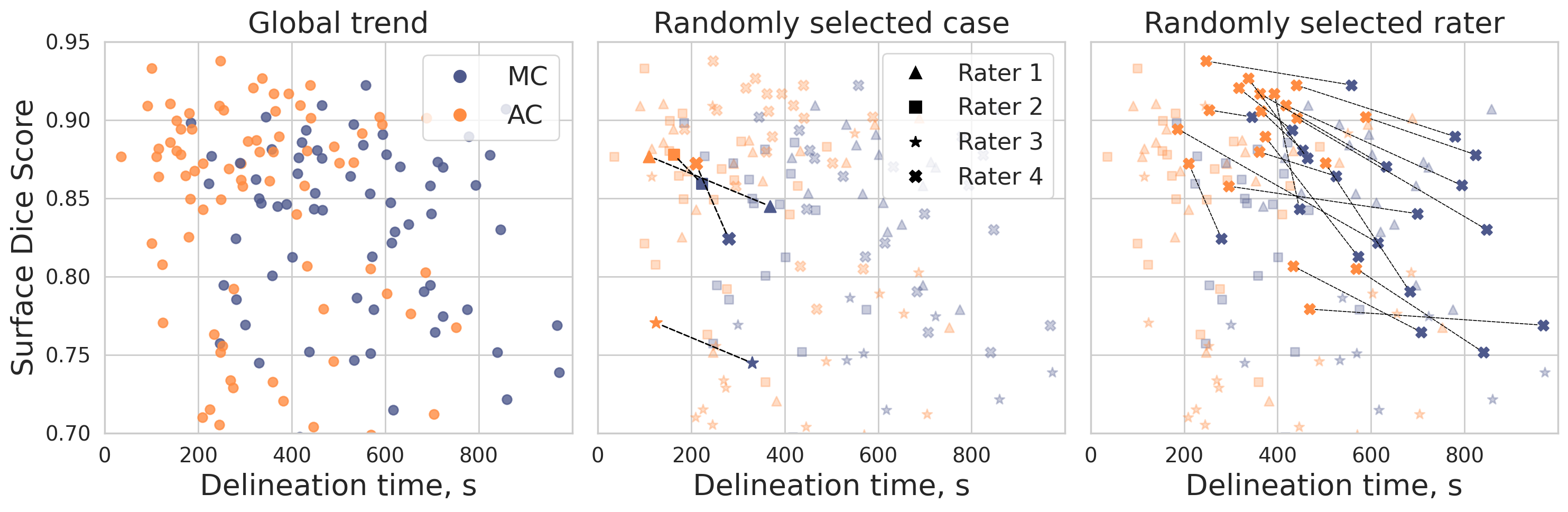}
    \caption{Delineation time \textit{vs.} inter-rater contouring agreement for every rater and every case. Contouring agreement is calculated as the average surface Dice Score (sDSC) per patient. Blue points correspond to manual contouring (\textit{MC}), and orange points correspond to assisted contouring (\textit{AC}). We restrict the time-axis to the maximum of $1000$ s and the sDSC-axis to the minimum of $0.7$; thereby, we leave some points outside the plot.}
    \label{fig:times_dices}
\end{figure*}

Finally, we show the simultaneous improvement of the inter-rater contouring agreement and the contouring time reduction (see Fig. \ref{fig:times_dices}). Since the contouring time is recorded case-wise, we use the average tumor-wise surface Dice Scores for every case. Here, Fig. \ref{fig:times_dices} shows the global trend of the CNN-initialized contouring to be less time-consuming and less subjective in terms of inter-rater variability.

\section{Discussion}
\label{sec:discussion}

Below we summarize our evaluation results, discuss the most important limitations of our study, and suggest the possible directions for future work.


Firstly, we extend most of the previous studies with the analysis of inter-rater detection variability. Our methodology suggests that the CNN assistance, on average, significantly reduces the probability of incorrect detection from $0.162$ to $0.085$ ($p = 0.036$). Although we have a significant improvement on average, the improvement for every single rater lacks enough evidence. The obtained results are statistically significant only for Rater $2$. However, a positive trend of the detection agreement improvement indicates that the results might be statistically confirmed with the larger prospective dataset in future work. Moreover, we highlight that the order of manual and assisted contouring techniques biases the evaluation of detection variability: $\Delta \hat{p}_{err} = 0.034$ in Group $1$ and $\Delta \hat{p}_{err} = 0.121$ in Group $2$. The latter might indicate that a one-week memory washout and even a six-week period, as in \cite{lu2021randomized}, are not enough for a rater to delineate the same image twice independently. We also present a qualitative analysis with the typical examples of detection errors in Fig. \ref{fig:fp} and Fig. \ref{fig:fn} as well as the analysis of the sizes of detection errors.



On the other hand, the analysis of the inter-rater contouring agreement is not affected by order of manual and assisted contouring techniques. We merge both groups and report an improvement of $0.026$ surface Dice Score ($p < 0.05$), and $0.020$ concordance index ($p < 0.05$). However, Rater $1$ does not achieve a statistically significant result. We note that Rater $1$ is the most experienced in the team; thus, we expect his contours to be the most accurate and receive minor changes from the CNN assistance. We also highlight the bias of the trained CNN to approximate the contouring style of Rater $1$. Here, approximately half of the development dataset on which the model is trained was annotated by Rater $1$. This message also resonates with \cite{wong2020comparing}, where the authors show that if a single rater annotates the dataset, the resulting models may not reduce the uncertainties in the tumor delineation.


Furthermore, the results suggest that the CNN assistance accelerates the contouring process from $1.6$ to $2.0$ times on average. Similar to the detection variability analysis, contouring time reduction significantly depends on the order of contouring techniques ($p = 0.019$). Therefore, the exact median relative speedup could not be given. The same bias also affects the median absolute time reduction; the CNN assistance accelerates the contouring process from $03$:$26$ to $04$:$53$ (\textit{mm:ss}). However, the results are statistically significant within both groups and for all raters, without exception.

\subsection{Conclusion}
\label{ssec:discussion:conclusion}

We have presented a clinical evaluation of a deep learning segmentation method for metastatic brain lesions in terms of (i) inter-rater detection agreement, (ii) inter-rater contouring agreement, and (iii) time savings. Firstly, we have shown that the deep learning assistance, on average, significantly reduces the ratio of detection disagreement two times. Secondly, our results have suggested that using CNN-assistance in multiple metastases segmentation significantly reduces inter-rater contouring variability in terms of both volumetric and surface metrics. Thirdly, we have demonstrated a significant delineation process acceleration up to two times. Finally, we have carefully outlined and discussed possible biases in the clinical evaluation of the Deep Learning model for medical image segmentation.





\bibliographystyle{IEEEtran}
\bibliography{bib_clinical,bib_dl,bib_related,bib_our}


\end{document}